\documentclass[]{spie}  

\usepackage{graphicx}
\usepackage[colorlinks=true, allcolors=blue,unicode]{hyperref}
\usepackage{amsmath,amsfonts,amssymb}
\usepackage{siunitx}
\usepackage[usenames,dvipsnames]{color}
\usepackage{colortbl}
\usepackage{soul}
\usepackage{microtype}
\usepackage{pdflscape}
\usepackage{tablefootnote}
\usepackage{upgreek}
\usepackage{wrapfig}
\usepackage{morefloats}
\usepackage{multirow}
\usepackage{booktabs}
\usepackage[flushleft]{threeparttable}
\usepackage[toc,page]{appendix}
\usepackage{longtable}
\usepackage{comment}
\newcommand{\Namakanui}{N\=amakanui }
\newcommand{\Uu}{$`$\=U$`$\=u }
\newcommand{\Aweoweo}{$`$\=Aweoweo }
\newcommand{\Alaihi}{$`$Ala$`$ihi }

\title{Commissioning of \Namakanui on the JCMT} 

\author{Izumi Mizuno\supit{a}, Per Friberg\supit{a}, Ryan Berthold\supit{a}, Harriet Parsons\supit{a}, Chih-Chiang Han\supit{b} , Alexis Acohido\supit{a}, Graham Bell\supit{a}, David Berry\supit{a}, Dan Bintley\supit{a}, Ming-Tang Chen\supit{b}, Alyssa Clark\supit{a}, Jamie Cookson\supit{a}, Vernon Demattos\supit{a}, Jessica Dempsey\supit{a}, Jason Fleck\supit{a}, Kuo-chieh Fu\supit{b}, Miriam Fuchs\supit{a}, Sarah Graves\supit{a}, Paul Ho\supit{a}, Sung-Po Hsu\supit{b}, YauDe Huang\supit{b}, Xue-Jian Jiang\supit{a}, Derek Kubo\supit{b}, John Kuroda\supit{a}, Shaoliang Li\supit{a}, Steve Mairs\supit{a}, Callie Matulonis\supit{a}, Mailani Neal\supit{a, d}, Neal Oliveira\supit{a}, Maren Purves\supit{a}, Mark Rawlings\supit{a}, Ya-Che Shih\supit{b}, Kevin Silva\supit{a}, Ed Sison\supit{a}, Patrice Smith\supit{a}, Ranjani Srinivasan\supit{b}, William Stahm\supit{a}, Alexandra J. Tetarenko\supit{a}, Pablo Torne\supit{a, c}, Craig Walther\supit{a}, Chao-Ching Wang\supit{b}, Ta-Shun Wei\supit{b}
\skiplinehalf
\supit{a}East Asian Observatory, 660 N. A'oh\={o}k\={u} Place, HI, USA 96720;\\
\supit{b}Academia Sinica, Institute of Astronomy \& Astrophysics P.O. Box 23-141, Taipei, Taiwan;\\
\supit{c}Instituto de Radioastronom\'ia Milim\'etrica (IRAM), Avda. Divina Pastora 7, Local 20, 18012 Granada, Spain;\\
\supit{d}New Mexico Institute of Mining and Technology, 801 Leroy Pl, Socorro, NM 87801, USA
}

\begin{document} 
\maketitle 
\begin{abstract}
\Namakanui is an instrument containing three inserts in an ALMA type Dewar. The three inserts are \Alaihi, \Uu and \Aweoweo operating around 86, 230 and 345GHz. The receiver is being commissioned on the JCMT. It will be used for both Single dish and VLBI observations. We will present commissioning results and the system.
\end{abstract}



\section{INTRODUCTION}
\label{sec:intro}  
The JCMT is the largest single dish telescope dedicated to sub-millimeter observing in the world. It has a 15-m primary dish and is often recognized by its iconic GoreTex membrane that enables it to observe in higher wind-speeds than would otherwise be possible. Located at 14,000' on Maunakea in Hawai`i, the high altitude combined with typically stable inversion layer trapping moisture at lower elevations, enables the JCMT to observe in some of the best sub-millimeter observing conditions in the Northern Hemisphere. The JCMT has a suite of instruments that enable the observatory to execute continuum observing (SCUBA-2 at 450 and \SI{850}{\micro\metre}\cite{Holland2013}), linear continuum polarization (POL-2, ancillary to SCUBA-2 for use at both 450 and 850 \SI{850}{\micro\metre}\cite{Friberg2018}), and heterodyne observations (HARP\cite{Buckle2009}, operating between 325 to 375\,GHz). The JCMT is now also home to the new \Namakanui receiver, that will be used both for single dish observations and also for VLBI observations such as the Event Horizon Telescope (EHT) and with the East Asia VLBI Network \cite{EAVN2018}.

In this paper we discuss the details of the \Namakanui receiver and its three inserts: \Alaihi, \Uu and \Aweoweo. Software to operate and monitor \Namakanui has been developed and we present the performance of \Namakanui and commissioning results from \Uu . Currently \Uu is available to the community for shared risk science observations, and we present some recent science observation taken with \Uu. 

\section{Overview of N\=amakanui}
The \Namakanui instrument is a spare receiver for the Greenland Telescope (GLT) \cite{Han2018,Bintley2018}, on loan to the East Asian Observatory for use on the JCMT courtesy of ASIAA. The \Namakanui contains three inserts: \Alaihi, \Uu and \Aweoweo operating around 86, 230 and 345GHz. 

One of the insert of \Alaihi is a new frequency range for JCMT, operating around 86\,GHz. The \Uu insert replaced the JCMT's RxA3 single polarization dual side-band receiver that was retired in June 2018 and operated between 212 and 274\,GHz. \Uu will cover most of the RXA3 operation frequency range, but with higher sensitivity by receiving dual polarizations. The \Aweoweo insert has an operational sky frequency range that overlaps much of the HARP operation range although with increased coverage at lower frequency than HARP. \Aweoweo will have a greater sensitivity than HARP due to its dual polarization capabilities, and lower receiver noise - and will outperform HARP for point source observations, although HARP will remain key for large scale mapping thanks to its multiple receptors\cite{Buckle2009}.

\begin{figure}
    \centering
    \includegraphics[width=0.8\linewidth]{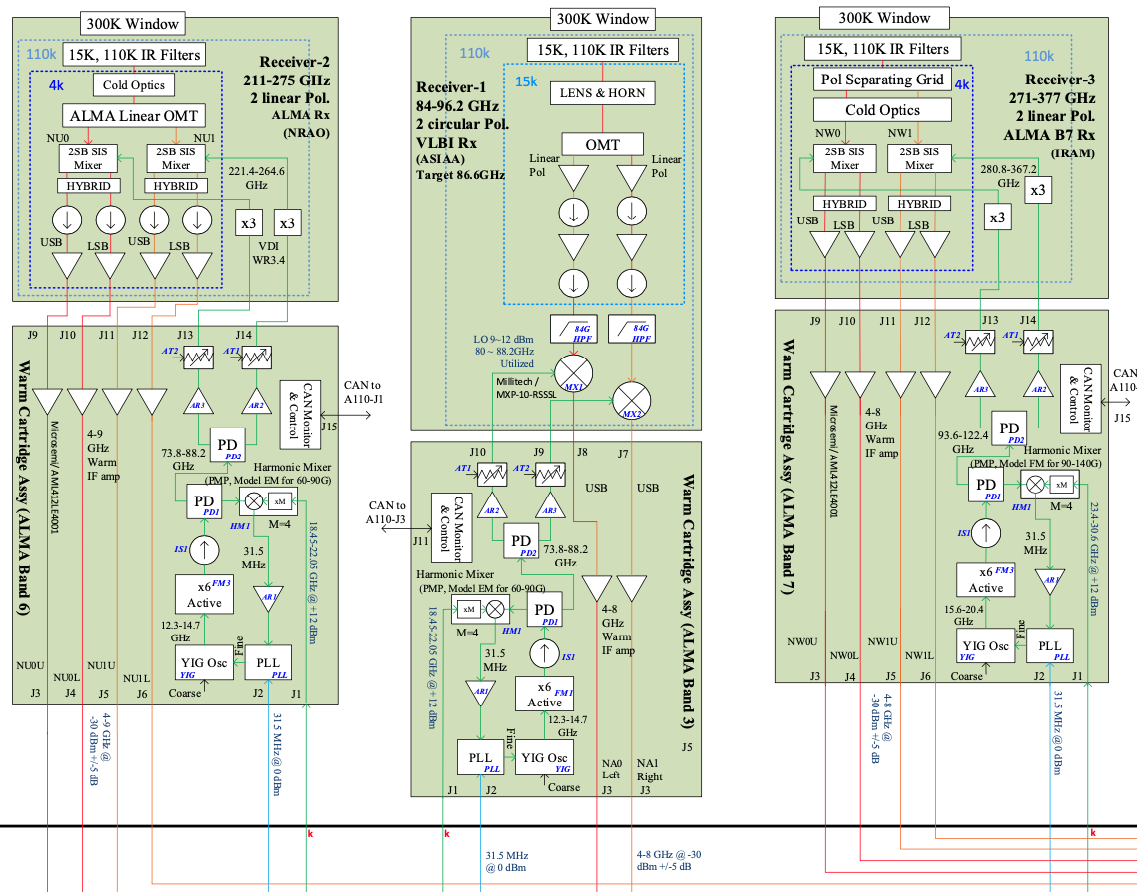}
    \caption{Block diagram for \Namakanui. Left: \Uu 230\,GHz, Middle: \Alaihi 86\,GHz Right: \Aweoweo 345\,GHz.}
    \label{fig:Namakaui_diagram}
\end{figure}

\begin{figure}
    \centering
    \includegraphics[width=0.8\linewidth]{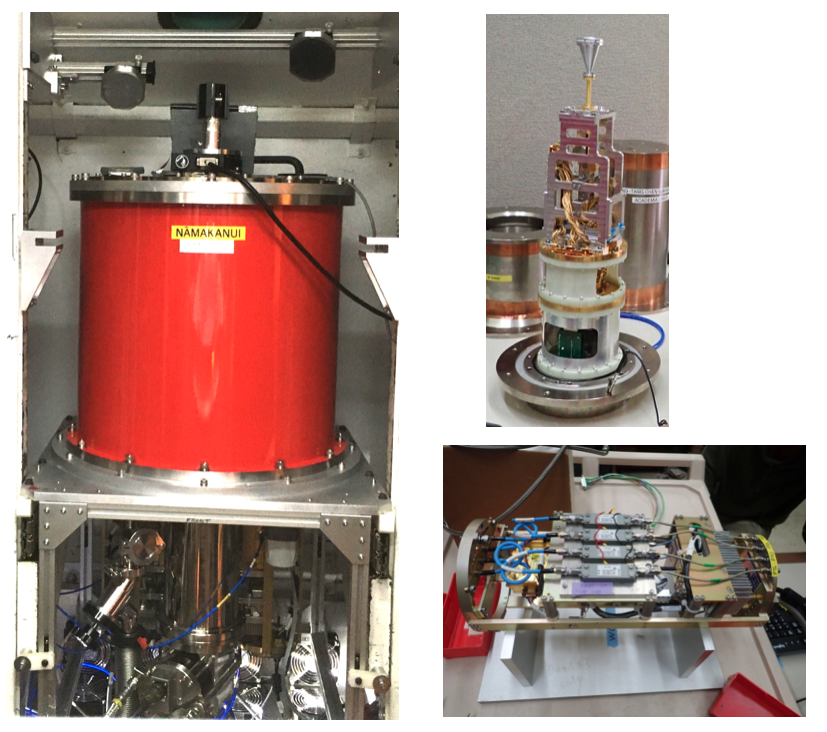}
    \caption{\Namakanui installed in cabin (left), with the cold cartridge assembly of \Alaihi (top right) and the warm cartridge assembly of \Aweoweo (bottom right)}
    \label{fig:Namakanuipic1}
\end{figure}

Each of the three \Namakanui inserts is composed of an ALMA (Atacama Large Millimeter/submillimeter Array) type cold cartridge and warm cartridge (e.g. Figure~\ref{fig:Namakanuipic1}). The cold cartridges are inserted in the Dewar and have mixers and low noise amplifiers. The warm cartridges are attached under their respective cold cartridges outside of the dewar and have signal oscillators for 1st LO signal and amplifier for IF signal.  Figure~\ref{fig:Namakanuipic1} is a picture of a cold, and warm cartridge, and Figure~\ref{fig:Namakaui_diagram} show block diagram for each inserts. All inserts controlled by a ALMA's electronics, FEMC (Front End Monitor and Control).

Although \Namakanui is a spare receiver for a near identical receiver in use at the GLT, \Namakanui has some differences. ASIAA upgraded the dewar design and differs tuning range for a insert of 230\,GHz (\Uu), and has different polarizations at 86\,GHz (\Alaihi). \Alaihi has the same design as the GLT 86\,GHz insert \cite{Han2018}, except for lacking a phase shifter. \Uu have same design as ALMA band 6 insert \cite{Ediss2004,Kerr2014} expect for LO wave guide and \Aweoweo have the same design as ALMA Band 7\cite{Mahieu2012} insert. 

A summary of the inserts is provided in Table \ref{tab:InsertSummary}. All cartridges receive dual linear polarization. We identify each polarization in this paper via the labels P0 and P1. Each insert has a separate optical path and so only one receiver/frequency range can be used at any one time.

\begin{table}[ht]
\caption{Summary of the three \Namakanui inserts. *LO frequency range of \Alaihi and \Aweoweo are preliminary values.} 
\label{tab:InsertSummary}
\begin{center}       
\begin{tabular}{|l|c|c|c|c|} 
\hline
\rule[-1ex]{0pt}{3.5ex}  Name & Receiver type & LO Frequency (GHz)* & Output IF\\
\hline
\rule[-1ex]{0pt}{3.5ex} \Alaihi & SSB  & 77.0--88.5 & 2IF (two pol., USB)\\
\hline
\rule[-1ex]{0pt}{3.5ex} \Uu & 2SB & 221--264.6 & 4IF (two pol., two sidebands) \\
\hline
\rule[-1ex]{0pt}{3.5ex} \Aweoweo & 2SB & 283--365 & 4IF(two pol., two sidebands) \\
\hline
\end{tabular}
\end{center}
\end{table} 


\Namakanui can be used with the spectrometer ACSIS (Auto Correlation Spectral Imaging System \cite{Dent-ACSIS-2000}), and the VLBI backend (R2DBE\cite{Casper2013}, and Mark6\cite{MIT2020}). 
The active insert's signals are selected by a switch, and the signals are divided, and fed to both backends. Please see paper 11453-142 of this conference for detail of the integrated system. 

\Namakanui was delivered to the facility in August 2019 and installed at the telescope in September 2019. First light was achieved early October 2019 with the 230\,GHz, Band 6 insert \Uu. 


\section{N\=amakanui: Optics}

\Namakanui is located in the Cassegrain receiver cabin which moves with the Antenna. Because all inserts are linear polarized, a quarter wave plate, one for each insert, is inserted during VLBI to covert from linear to circular polarization.  An ambient load on a rotation stage can be inserted in front of each insert.


The \Namakanui Dewar optics is designed for the GLT, which is one of the ALMA prototype dishes. Hence, the Dewar is intended to be located on the telescope optical axis with the top surface at the Cassegrain focal plane. The JCMT has a steerable tertiary mirror unit (TMU) on the telescope optical axis before the Cassegrain focal plane, which reflects the incoming beam to one of several receiver bays located in the Cassegrain cabin surrounding the TMU. The TMU can also reflect the beam to one of the Nasmyth platforms. In addition the optics in the \Namakanui Dewar match the F/8 optics used by the ALMA dishes and not the F/12 optics used by the JCMT dish. While it would be possible to redesign the optics in the \Namakanui Dewar, that was not an option as \Namakanui is spare receiver for the GLT and must remain ready for use at the GLT. 

To couple the \Namakanui optics to the JCMT telescope the Dewar was mounted with its optical axis vertical in one of the receiver bays (vertical is referring to the location when the JCMT is pointing to the zenith, see \ref{fig:Namakanuipic1}). The Dewar optical axis is then parallel to telescope optical axis. The steerable TMU is used to reflect the incoming signal to a point above the window on the \Namakanui Dewar. A powered coupling mirror is used to reflect the signal into the Dewar. By selecting the height above the Dewar and the power of the mirror the focal planes and F ratios can be matched. There is one coupling mirror for each insert and the angle of incident is about 35 degrees. The main design issue was the space limitations to fit in three mirrors as well as calibration loads above the dewar. A more elaborate optical design using more than one mirror per insert was ruled out due to the mechanical complexity. Even when using one coupling mirror per insert, the Cassegrain ceiling had to be removed to fit in the 86\,GHz mirror between two major members of the dish backup structure. 
Coupling mirrors for \Uu and \Aweoweo are installed; alignment of the coupling mirrors was done by using thermal loads, and the TMU was aligned using the Sun.

\section{N\=amakanui: Tuning}
Tuning a \Namakanui receiver cartridge to a particular LO frequency is done in several steps: 1) performing initial setup; 2) finding a PLL (phase-locked loop) lock; 3) adjusting the PLL control voltage; 4) setting mixer parameters, and 5) servoing the power amplifiers.  For additional details, refer to the ALMA document\footnote{\url{https://ictwiki.alma.cl/twiki/pub/Main/FrontEndToControlSoftware/FEND-40.00.00.00-089-A-MAN.pdf}}.

Initial setup consists of setting the IF switch to the desired receiver insert and adjusting the external reference signal generator.  The reference signal is coupled to the WCA (Warm Cartridge Assembly) PLL via a 4X harmonic mixer, so the desired frequency is (LO/COLDMULT + FLOOG)/4, where COLDMULT is the cold-cartridge LO multiplication and FLOOG is the first LO offset generator frequency, 31.5 MHz.  The reference signal output power is set via a lookup table to keep it in the optimal range for the PLL.

Next, the YIG tuned oscillator (YTO) is adjusted until the phase-locked loop acquires an initial lock.  The target YTO frequency is LO/MULT, where MULT is total LO multiplication in both cartridges, COLDMULT*WARMMULT where WARMMULT is WCA LO multiplication factor.  Coarse YTO adjustment is done via a 12-bit value; the initial value is computed using the min/max YTO frequencies from a configuration file.  The YTO counts are then set above and below this initial value in increasing steps until a lock is found.

Fine adjustment of the YTO is done via a control voltage between +-10V set by the PLL.  To ensure a stable lock, the YTO coarse counts are adjusted until this control voltage is close to 0V.  First, the coarse counts are roughly adjusted using a known estimate of the YTO frequency vs voltage relationship.  Then the YTO counts are single-stepped toward 0V until the PLL control voltage changes sign.

Once the PLL is locked and adjusted, lookup tables are used to set SIS mixer parameters to their desired values for the current LO frequency.  These parameters are magnet currents, low-noise amplifier drain voltages/currents, and mixer bias voltages.  The lookup tables are typically defined across the LO range with entries at 4\,GHz spacing, using linear interpolation between points.

\begin{figure}
  \centering
  \includegraphics[width=0.7\linewidth]{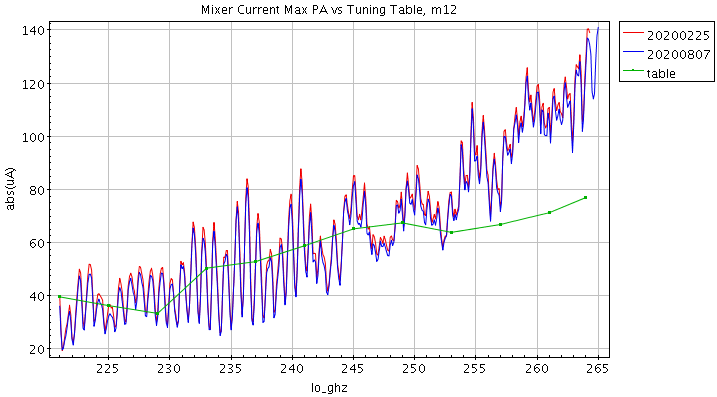}
  \caption{Mixer current as a function of LO frequency for one of the two P1 mixers. Blue and red lines relate to the date of measurements, red 2020/02/25, blue 2020/08/07. The green line indicates optimal mixer current. From the green line we see that the optimal mixer current cannot be reached at all frequencies} 
  \label{Lo_vs_mixcurrent}
\end{figure}

Finally, the warm cartridge LO power amplifiers are adjusted to achieve the desired SIS mixer current.  Initial PA (Power Amplifier) gate and drain voltages are taken from a lookup table, then the drain voltage is increased or decreased as needed.  At certain frequencies, the optimal SIS mixer current cannot be reached\ref{Lo_vs_mixcurrent}. The points of low mixer current correspond to spikes in the T\textsubscript{rx} plot seen in Figure~\ref{Uu_LOvstrx}.

Note that during a typical observation, the full tuning procedure is performed only once, at the start.  For small Doppler corrections during the course of an observation, only the PLL is relocked, while the SIS/PA parameters are left at their previous values.  This is done to avoid power jumps due to the digital nature of the control system.

\section{N\=amakanui: Trx}

The sensitivity of a heterodyne instrument is given by the receiver temperature (T\textsubscript{rx}) combined with beam efficiency. The measured T\textsubscript{rx} for all three inserts are provided. 

\subsection{T\textsubscript{rx} value of \texorpdfstring{\Uu}{}}
\begin{figure}
    \centering
    \includegraphics[width=0.9\linewidth]{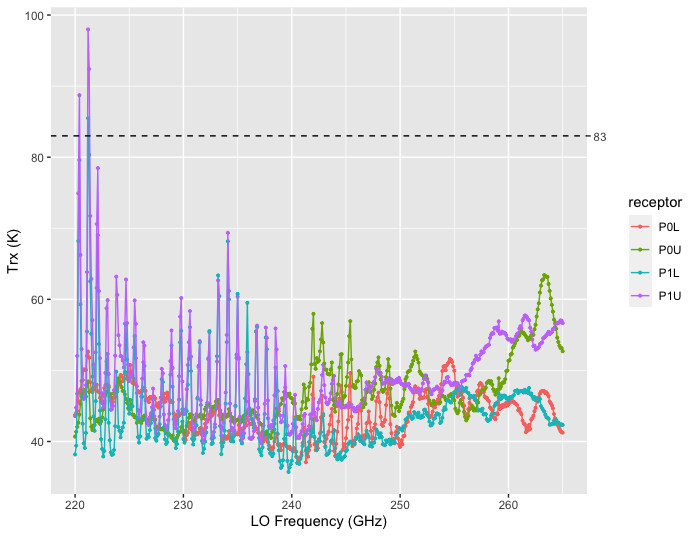}
    \caption{T\textsubscript{rx} as a function of LO observed with an IF of 6\,GHz for \Uu. Points are averaged over 1\,GHz bandwidth. P0 and P1 indicate polarization, with U and L indicating USB and LSB. Dashed line at 83K indicates the ALMA specification for T\textsubscript{rx} across the band.}
    \label{Uu_LOvstrx}
\end{figure}

\begin{figure}
    \centering
    \includegraphics[width=0.9\linewidth]{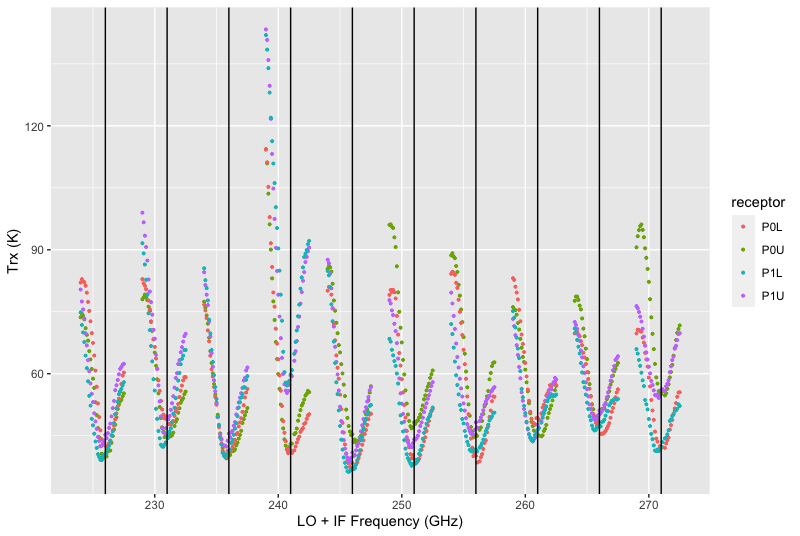}
    \caption{T\textsubscript{rx} as a function of LO and IF frequency for \Uu. Horizontal axis indicate addition of LO and IF frequency. Vertical axis show T\textsubscript{rx} value averaged over 1GHz bandwidth. LO frequencies are sampled from 220 to 265\,GHz with 5\,GHz step. IF frequency are sampled from 4 to 7.5\,GHz with 100 MHz step. Black vertical lines indicate 6GHz IF frequency points. P0 and P1 indicate polarization, and U and L indicate USB and LSB.}
    \label{Uu_LOvstrx_IF}
\end{figure}

The T\textsubscript{rx} across the \Uu tuning range is provided in Figure~\ref{Uu_LOvstrx}. It was measured using ACSIS at an IF of 6\,\,GHz IF and averaged over 1GHz bandwidth. Measurements were spaced by 100 MHz steps from 220\,GHz to 265\,GHz LO frequency. The T\textsubscript{rx} values are found to be between 35--98K. This places 99.8\% of the T\textsubscript{rx} values lower than the ALMA specification value of 83K\footnote{The T\textsubscript{rx} specification requires that 80\% of T\textsubscript{rx} values is lower than the value at ALMA\cite{Warmels2020}}. Figure~\ref{Uu_LOvstrx_IF} shows the relationship between IF and T\textsubscript{rx} for multiple LO frequencies.  IF frequencies are taken in steps of 100\,MHz ranging from 4 to 7.5\,GHz. It is found that the T\textsubscript{rx} reaches its lowest value with an IF around 6\,GHz across the whole LO frequency range.

As shown in Figure~\ref{Uu_LOvstrx}, T\textsubscript{rx} has narrow spiky features in both polarizations and sidebands. P1 has spiky features lower than 245\,GHz. P0 has the feature between 240 and 250\,GHz. At the regions where spikes exist we see issues of lack of LO pumping power. It was investigated that power decreases at either a connection between the warm and cold cartridges or in the cold cartridge. The frequency span between peak and bottom is about 300-600 MHz, and observers could avoid the spike by shifting the observation frequency. The JCMT Observing Tool (OT) has a newly added T\textsubscript{rx} graph to avoid the T\textsubscript{rx} peaks (see OT instruction\footnote{\url{https://www.eaobservatory.org/JCMT/observing-tool/het_example.html\#het_adv}}
 for details). 

\subsection{T\textsubscript{rx} of \texorpdfstring{\Aweoweo}{}}
\begin{figure}
   \centering
    \includegraphics[width=0.9\linewidth]{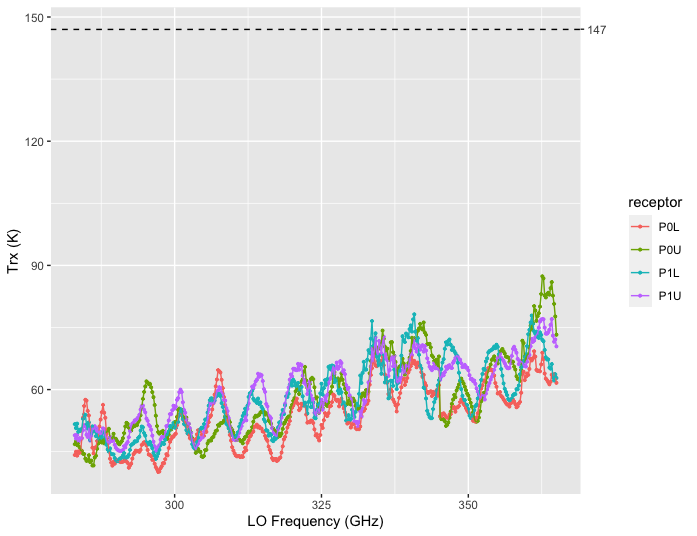}
    \caption{T\textsubscript{rx} as a function of LO observed with an IF of 6\,GHz) for \Aweoweo. Points are averaged over 1GHz bandwidth. P0 and P1 indicate polarization, with U and L indicating USB and LSB. Dashed line at 147K indicates the ALMA specification for T\textsubscript{rx} across the band.}
    \label{Aweoweo_LOvstrx}
\end{figure}

The relationship between T\textsubscript{rx} and LO frequency for the \Aweoweo insert is provided in Figure~\ref{Aweoweo_LOvstrx}. Once again the T\textsubscript{rx} was measured using ACSIS at an IF of 6\,GHz and averaged over 1GHz bandwidth and sampled with 100 MHz steps this time from 283 to 365\,GHz LO frequency. The T\textsubscript{rx} is found to be between 40 and 87K. This time all frequencies fall below the 80\% ALMA specification requirement of 147K\cite{Warmels2020}.

\subsection{T\textsubscript{rx}  of \texorpdfstring{\Alaihi}{}}
\begin{figure}
    \centering
    \includegraphics[width=0.7\linewidth]{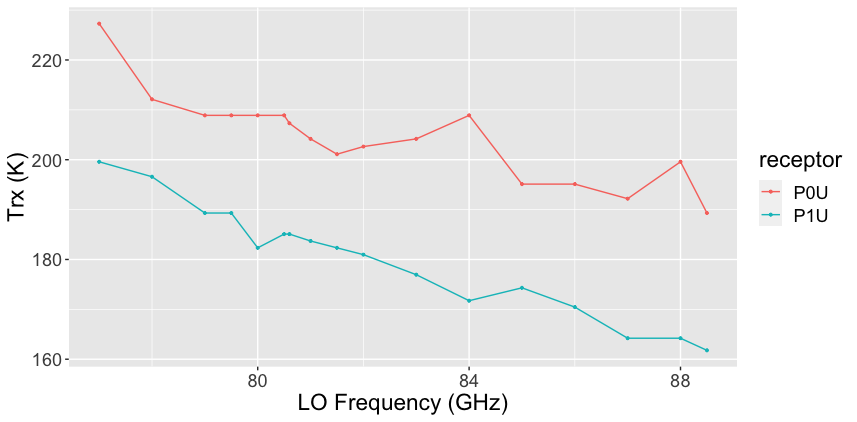}
    \caption{T\textsubscript{rx} as a function of LO observed with an IF of 6\,GHz) for \Alaihi. Points are averaged over 1GHz bandwidth. P0 and P1 indicate polarization, with U and L indicating USB and LSB. }
    \label{Alaihi_LOvstrx}
\end{figure}

The relationship between T\textsubscript{rx} and LO for \Alaihi averaged over the whole IF frequency range is shown in Figure~\ref{Alaihi_LOvstrx}. The T\textsubscript{rx} value is 161--227 K, and it is higher than expected values which are the typical T\textsubscript{rx} (60--80K\cite{Han2018}) of 86GHz insert at GLT. 

\section{N\=amakanui: Phase stability}
\begin{figure}
  \centering
    \includegraphics[width=0.9\linewidth]{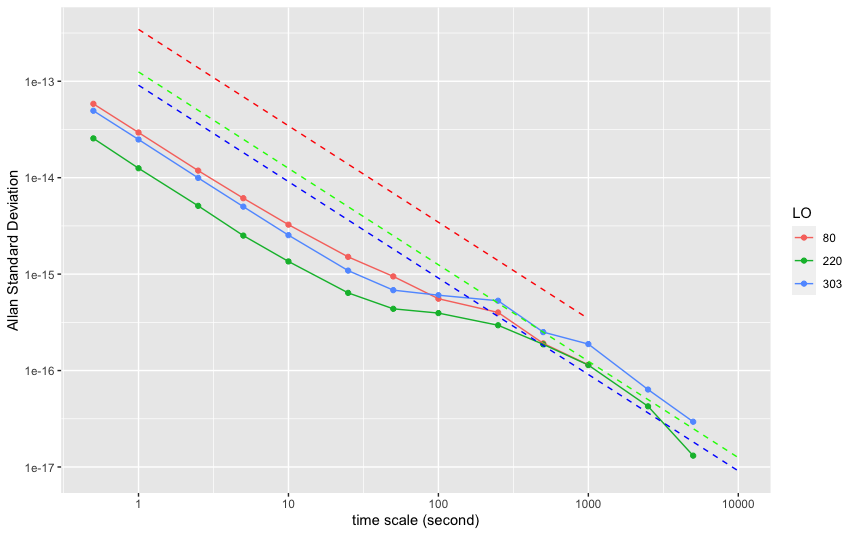}
  \caption{Allan standard deviation - a  measure  of  frequency  stability for each of the three inserts. Colors indicate LO frequencies.  LO frequencies of 80\,GHz, 220\,GHz, 330\,GHz are produced by \Alaihi, \Uu, and \Aweoweo respectively. The dashed lines indicates the Allan standard deviation causing a 1\% coherence loss.}
  \label{phaseAV}
\end{figure}

Phase stability of all three inserts is measured using an artificial monochromatic radio signal. The monochromatic signal is injected to an insert, and the phase difference between output signal from each insert and a signal produced by the same reference signal as the injected signal is measured by a vector volt meter. 
The reference signal fluctuation is cancelled out, and the Allan deviation of the phase describes the phase stability of the inserts. The measured Allan deviation, and the Allan deviation which would cause a 1\% coherence loss \cite{2017Thompson}, are shown in Figure~\ref{phaseAV}. All three inserts cause less than 1\% coherence loss by 100 seconds time scale. Typical coherence time in sub-millimeter observations is restricted by atmospheric fluctuation, and is less than 100 seconds.
Additionally stability of the integrated VLBI observation system including another critical factor for VLBI, LO phase noise with \Uu is confirmed as applicable for VLBI by detecting fringes with the SMA (The Submillimeter Array) and the GLT. 

\section{Commissioning \texorpdfstring{$`$\=U$`$\=u (230\,GHz insert)}{}}

The \Uu insert is the first of the three \Namakanui inserts that the observatory is currently commissioning. First light with \Uu was achieved on 2019/10/04.

A minor setback in the commissioning of \Uu was a problematic mixer of the P0 2SB assembly. A chiller fault caused \Namakanui to partially warm up in early November 2019 and after getting it back to operational temperature P0 was found to be dead with a flat line on the IV curve indicating no mixer current. This was to be a temporary failure, as a second unrelated chiller fault later the same month lead to a full warm up and cool down cycle on \Namakanui which appeared to revive the dead P0 mixer. The mixer that temporarily failed was the mixer that was known to be problematic with a soft IV curve. After discussion a spare mixer assembly was installed in early February with the P0 mixer assembly performing well ever since. The commissioning results shown below are taken after the mixer replacement.
\subsection{\texorpdfstring{$`$\=U$`$\=u performance: Pointing and Beam size}{}}
A pointing model has been developed for \Uu and we find the pointing accuracy to be within 3 arcseconds in both elevation and azimuth. The beam size and pointing are identical for both polarization. The FWHM (Full Width at Half Maximum) of the beam as a function of frequency is found to be described by:

\begin{equation}
    FWHM_{arcsecond} = 5.73 + \frac{14.68\times230}{f_{GHz}}
\end{equation}

This provides a FWHM of 20" at the commonly used \Uu frequency of CO (2-1), 230.538\,GHz.




\subsection{\texorpdfstring{$`$\=U$`$\=u performance: Calibration}{}}

The calibration of heterodyne data can be considered a two step process. The initial calibration of the data using system temperature (T$_{sys}$) to yield antenna temperature T$_{A}^{*}$, and the subsequent calibration of data from T$_{A}^{*}$ to main beam temperature (most appropriate for observations of point like sources) or radiation temperature (most appropriate for sources filling the beam). In this section we focus on the calibration to antenna temperature using T$_{sys}$.

To obtain the system temperature we apply chopper wheel calibration using blank sky, an ambient load and a measurement of the atmospheric opacity during the observation. Opacity at 225\,GHz is taken from the 186\,GHz water vapor radiometer at JCMT, and applied to the whole \Uu operation frequency range with the assumption that the opacity does not significantly vary by frequency. 


To check the calibration accuracy, we carried out a number of observations towards standard sources and lines (both USB and LSB) spanning a range of optical depths.

\begin{figure}
    \centering
    \includegraphics[width=0.9\linewidth]{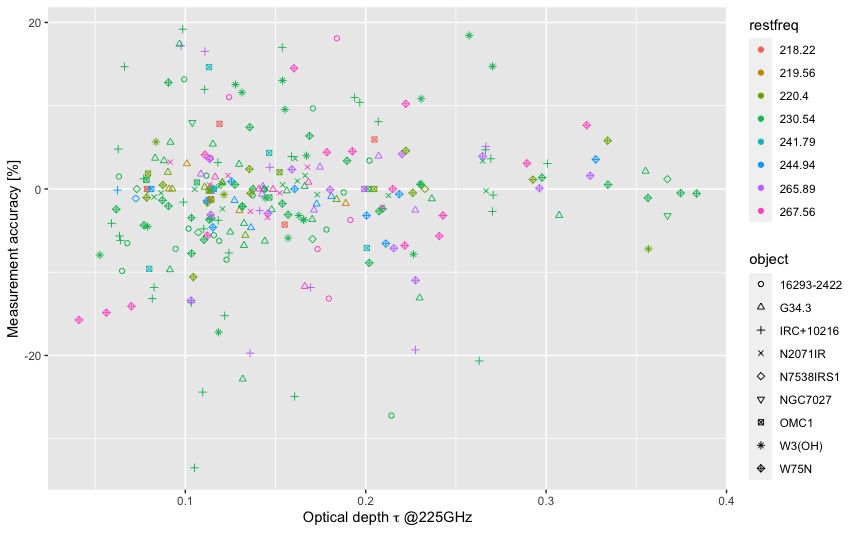}
    \caption{Consistency check of peak temperature across varying optical depths, found from observations of JCMT standards. Measurement accuracy here means the peak intensity from a single observation compared with the median peak intensity for that source. Point shapes and colors indicate object and rest frequency as in the legend. Data were taken between May and November 2020, using an IF of 6\,GHz. Lines $<$250\,GHz were obtained in LSB. Lines $>$230\,GHz are obtained in USB. No strong relationship with optical depth was found, indicating our T$_{sys}$ measurements are reasonable.}
    \label{tauvsdiff}
\end{figure}

We find the standard deviation of the difference (peak intensity from each observation compared to the median value for each source) is 7.6\%, and 98\% data is within $\pm 20\%$. There is no significant systematic difference by optical depth (see Figure~\ref{tauvsdiff}). 


\subsection{\texorpdfstring{$`$\=U$`$\=u performance: Leakage}{}}
The leakage of signal from one sideband into another sideband is well known\cite{Finger2013}. We measured the ratio of leakage by observing strong calibrator lines and dividing the line intensity in the desired sideband by the leaked intensity in the other sideband. We investigated the leakage ratio as a function of both LO frequency and also as a function of IF. 

\begin{figure}
    \centering
    \includegraphics[width=0.8\linewidth]{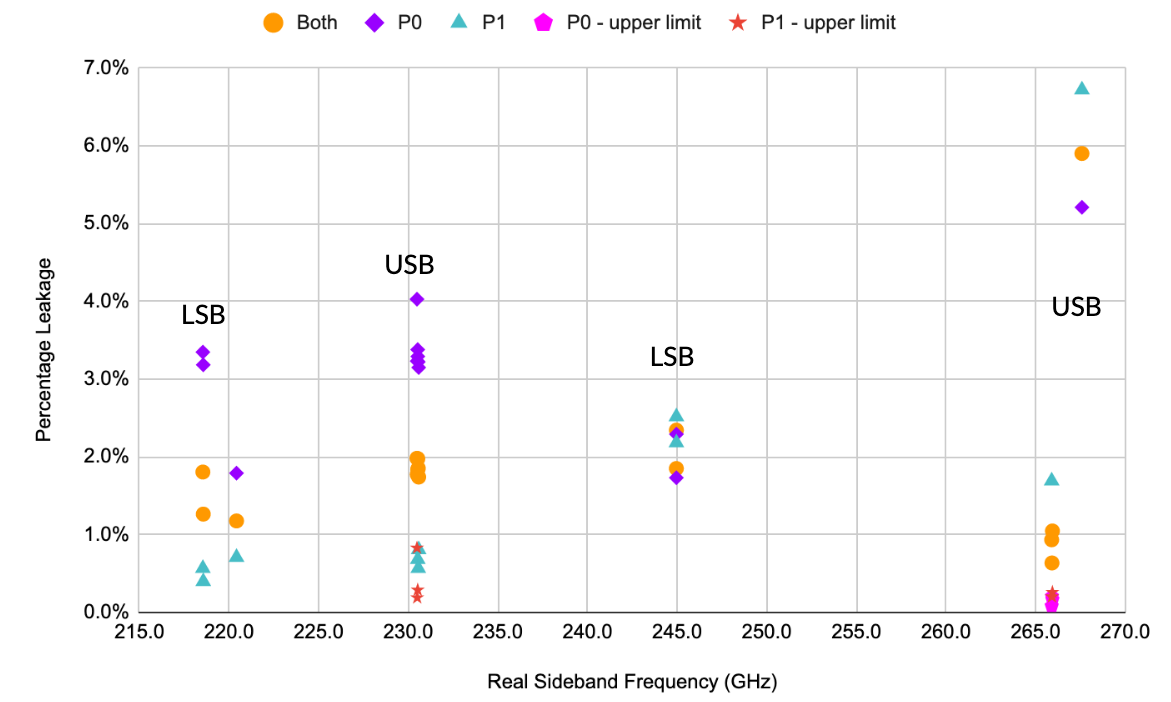}
    \caption{Leakage from signal sideband into image sideband as a function of frequency.  Below 235\,GHz the leakage is greater in P0 and frequencies above 235\,GHz leakage in greater in P1 independent of the sideband used.}
    \label{leakage-RF}
\end{figure}

\begin{figure}
    \centering
    \includegraphics[width=0.9\linewidth]{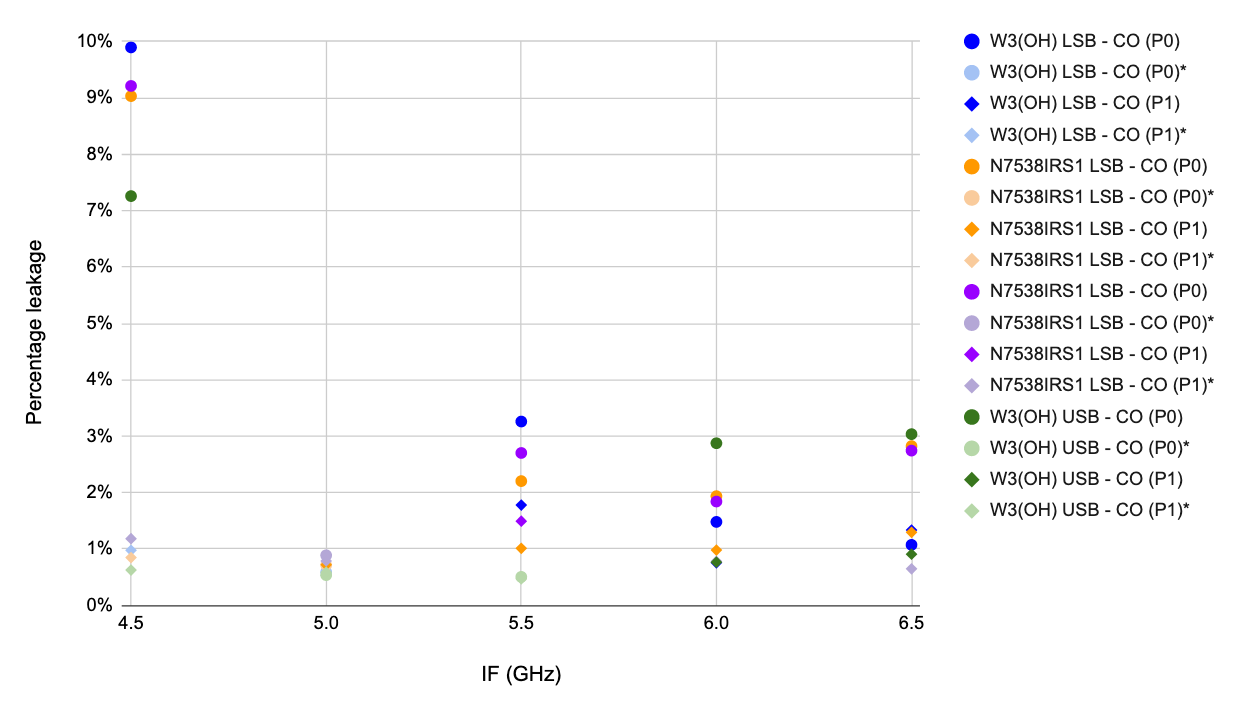}
    \caption{Leakage from signal sideband
    into image sideband as a function of IF. A * appended to the legend caption indicates upper estimate based on the noise. On average leakage across the IF range is $<$3\%, with the exception of P0 at an IF of 4.5\,GHz.}
    \label{leakage-IF}
\end{figure}

Across frequency space leakage was on average $\sim$3\%. Below 235\,GHz the leakage is greater in P0 and frequencies above 235\,GHz leakage in greater in P1 independent of the sideband used (see Figure~\ref{leakage-RF}). As a function of IF we found on average leakage across the IF range to be $<$3\% in both P0 and P1, with the exception of P0 at an IF of 4.5\,GHz (see Figure~\ref{leakage-IF}) although this work needs to be expanded to check the trend as a function across LO frequencies.

\subsection{\texorpdfstring{$`$\=U$`$\=u performance: Sideband comparison}{}}
To check the spectral sideband consistency and difference of \Uu, we observed CO (2-1) at a frequency of 230.538\,GHz, and CS (5-4) at a frequency of 244.936\,GHz towards eight JCMT standard sources\footnote{\label{EAO_calsource}\url{https://www.eaobservatory.org/jcmt/instrumentation/heterodyne/calibration/}}. The CO (2-1) and CS (5-4) lines were selected as the molecular lines can be placed in both LSB and USB.

Both P0 and P1 were considered separately when evaluating the sideband performance. It was found that peaks in P1 are generally slightly higher than that in P0, but only on a level of less than 5\% (see Fig 1-4 and Table \ref{table:sideband}).

\begin{table}
\caption{Median of the peak ratios. From the ratios one can tell how much difference between different sidebands or polarizations.
}
\begin{center}
\begin{tabular}{|c|c|c|c|c|}
\hline
  & Frequency &  \multicolumn{1}{c|}{LSB (median)} &  \multicolumn{1}{c|}{USB (median)} & \multirow{2}{6em}{USB/LSB using ``both"} \\
  & (GHz)  & P1/P0 & P1/P0 &  \\
\hline
CO (2-1) & 230.538 & 5.19\% & 1.80\% & 3.69\%  \\
CS (5-4) & 244.936 & 2.75\% & 4.45\% & 3.82\%  \\
\hline
\end{tabular}
\end{center}
\label{table:sideband}
\end{table}


\subsection{\texorpdfstring{$`$\=U$`$\=u performance: polarization comparison}{}}
\begin{figure}
  \centering
  \includegraphics[width=0.8\linewidth]{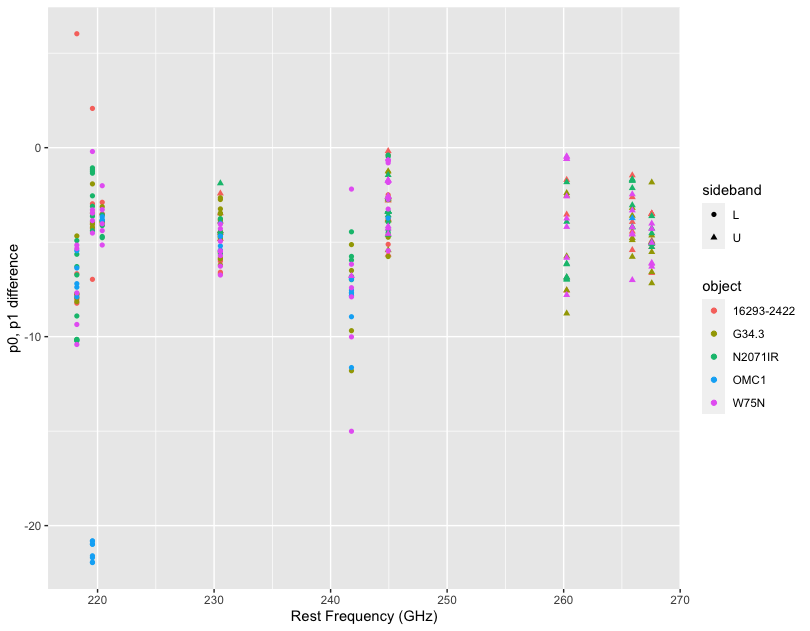}
  \includegraphics[width=0.5\linewidth]{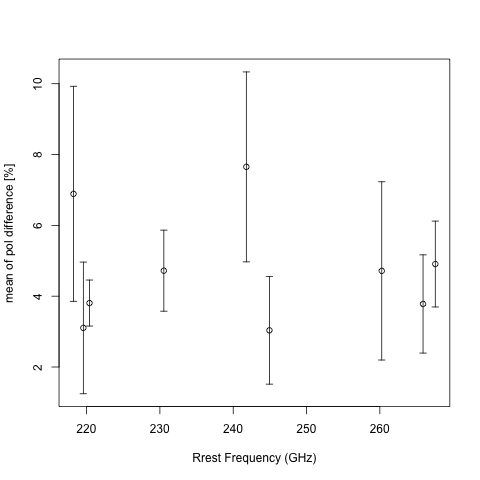}
  \caption{Relationship between rest frequency and polarization difference. Top: Polarization difference for JCMT standard source observations. Point shapes and colors indicate sideband and object as in the legend. Points with values $>0$ indicate P0 is higher. Points with values $<$0 indicate P1 is higher.} 
  \label{Poldiffvsrestfreq}
\end{figure}

To check the consistency and difference between P0 and P1 for \Uu, we compared peak antenna temperature between the two polarizations (see Figure~\ref{Poldiffvsrestfreq}). We used JCMT standard source$^{\footnotemark[0]{\ref{EAO_calsource}}}$  observations obtained with 2048 channels in 1000MHz band width, with an 6\,GHz center IF frequency. Peak antenna temperatures ranged between 1.8K and 91K.
The polarization difference was estimated by $(P0_{peak}-P1_{peak})\times200/(P1_{peak}+P0_{peak})$ for each observation, and  we estimated the median and standard deviation of the difference for each rest frequency, removing the OMC1 observations at 219\,GHz (see Figure~\ref{Poldiffvsrestfreq}). The mean difference is P1 is higher than P0 by 3\%--8\% across most of the \Uu frequency range.

In the case of the observations of OMC1 at 219\,GHz, P1 is 20\% larger than P0. The line is not intrinsically polarized. The line is observed at the edge of the 1\,GHz frequency band, and such larger difference was found when the line is not observed around center of bandwidth (see section \ref{sec:position_in_band} for details). 

\subsection{\texorpdfstring{$`$\=U$`$\=u performance: Signal variation within observation band}{}}
\label{sec:position_in_band}

\begin{figure}
    \centering
    \includegraphics[width=0.9\linewidth]{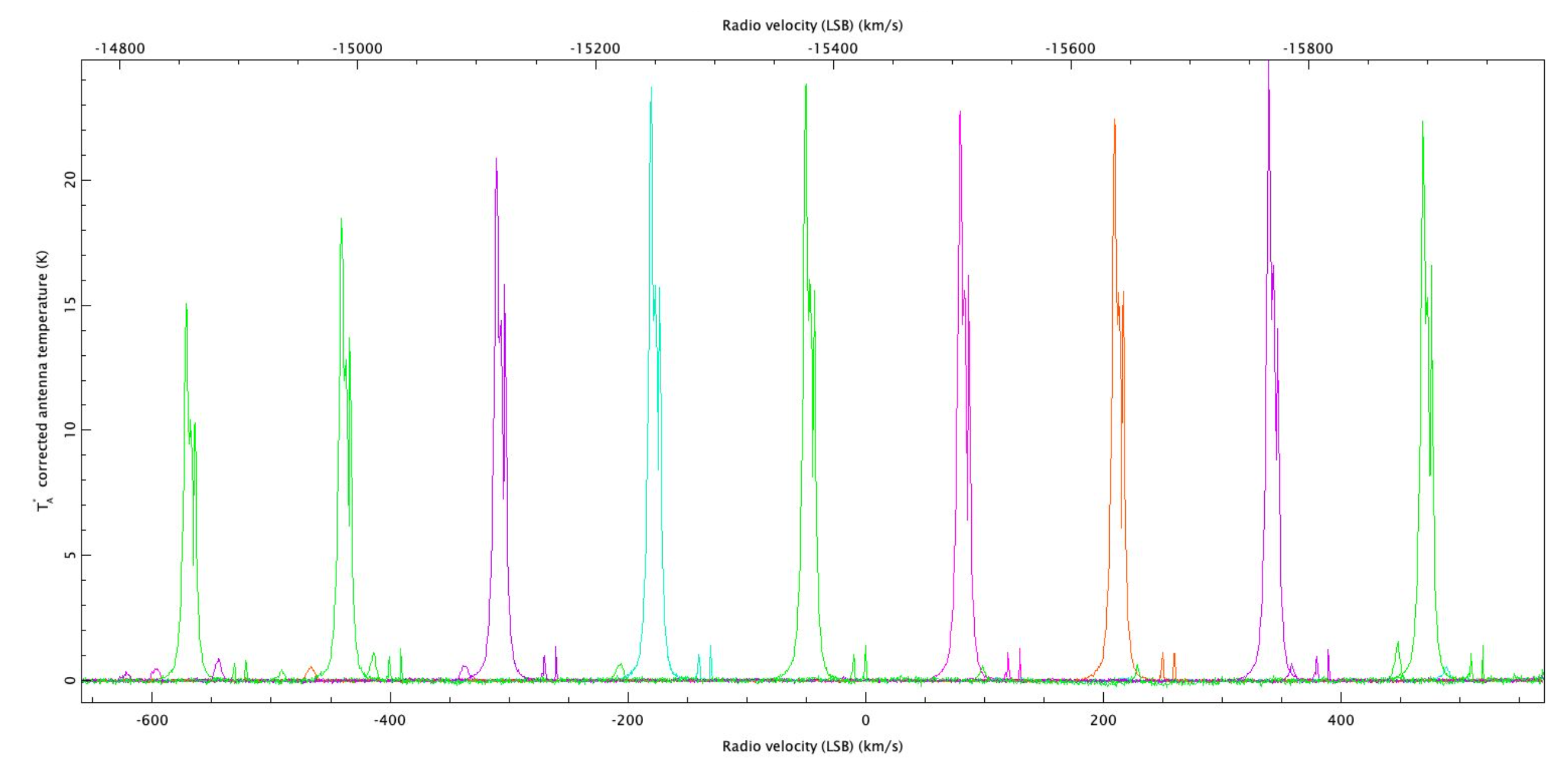}
    \caption{A series of 9 near identical observations of W3(OH) in CO (2-1) observed in LSB with a central IF of 6GHz with shifted LO  (observations taken from P0 LSB). It can be seen that position within the band affects the measured peak intensity.}
    \label{fig:spectra-shifted-in-band}
\end{figure}

During commissioning it was noted that the peak intensity appears to attenuate at LO or IF frequencies that puts a strong line of interest (e.g. CO (2-1)) towards the edge of the band. For P0 in LSB this is shown in Figure~\ref{fig:spectra-shifted-in-band}. 
This effect is most startling in LSB observations - with an attenuation of up to 30\% and is greatest in P0 compared to P1. Work is ongoing to determine the cause of the issue but currently it is believe that this may be caused from an ACSIS power issues.

\subsection{Baseline issue - P cygni profile}
\begin{figure}
    \centering
    \includegraphics[width=0.7\linewidth]{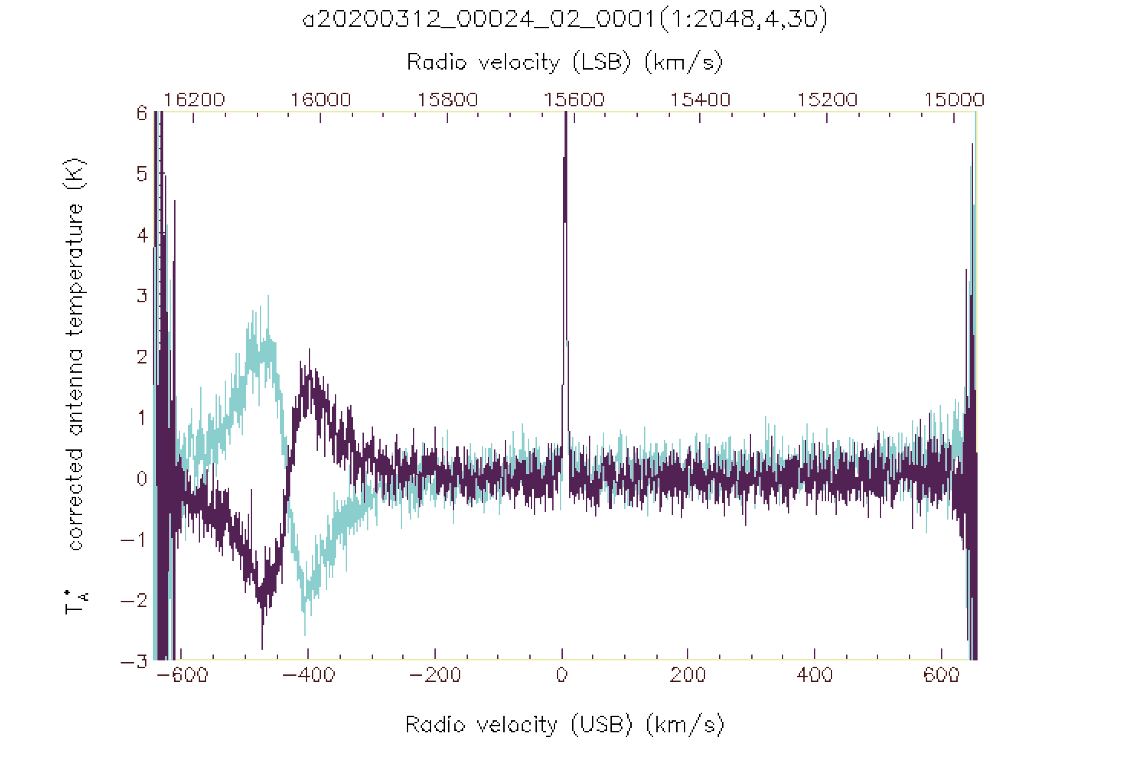}
    \caption{Two spectra are two continuous sub scan spectra in an observation. The artificial P cygni profile clearly inverted between continuous sub scan when telescope went off to the off position and then back. Such sudden jumps have only been found when going off to the off position}
    \label{fig:p_cygni}
\end{figure}

We have some undesired baseline feature due to atmospheric and system variation for some observations. Artificial P cygni profiles were observed in P1 USB in the edge of baseline in 1GHz band spectra. It is related to the telescope motion, and the shape varies during an observation with telescope motion, as shown in Figure~\ref{fig:p_cygni}.

\subsection{Baseline issue - ozone line}
\begin{figure}[!htbp]
  \centering
  \includegraphics[width=0.48\linewidth]{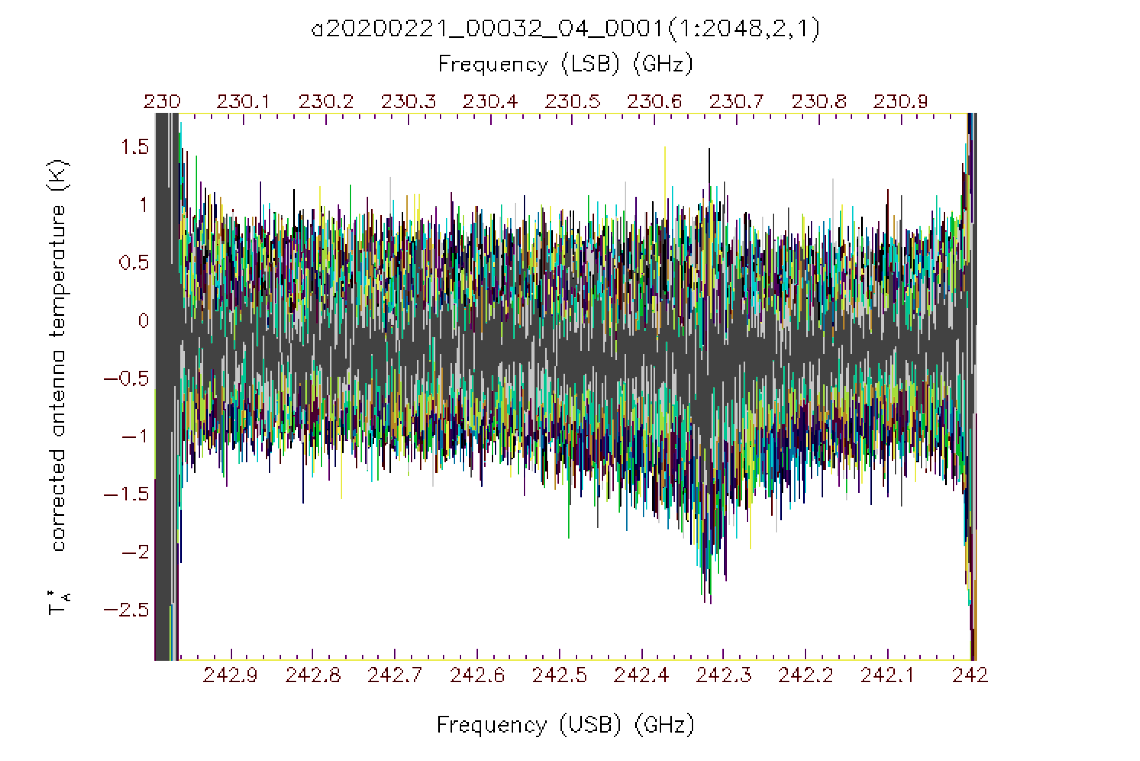}
  \includegraphics[width=0.48\linewidth]{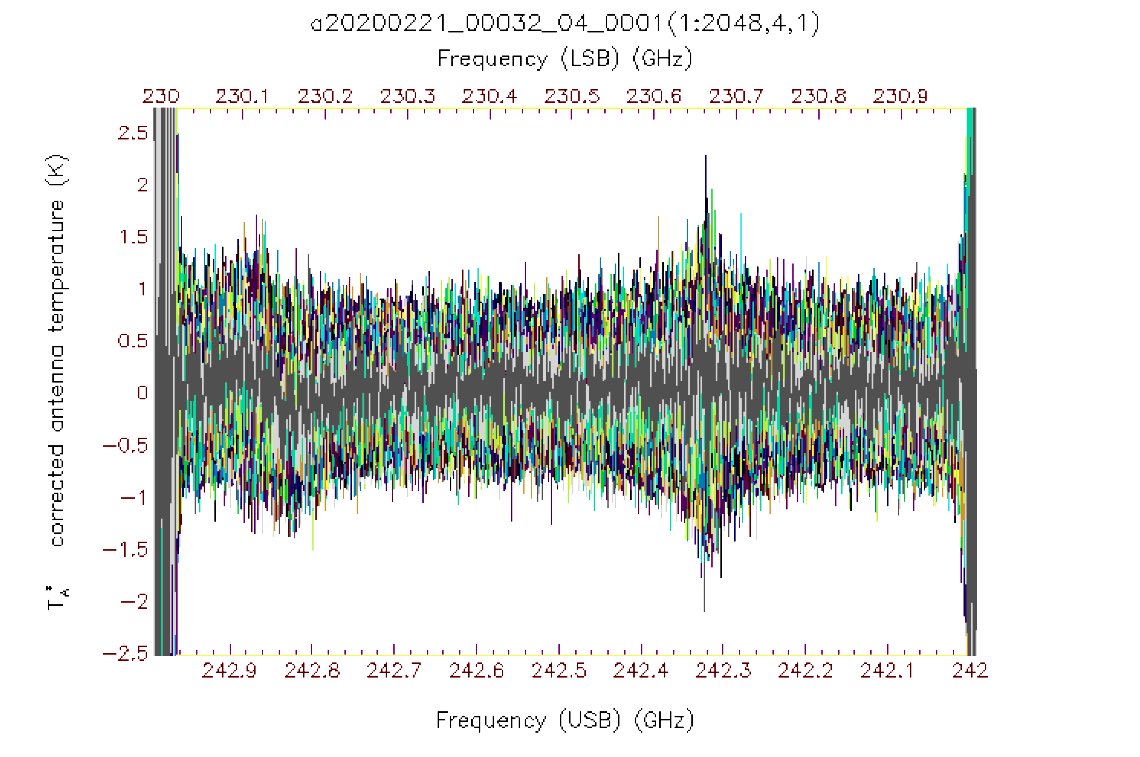}
  \caption{Spectra reduced by each sub scan for P0 (left), and P1 (right). Different color indicates different sub scan. Ozone line was found in 242.32\,GHz, and it goes up and down during the observation.}
  \label{fig:ozone}
\end{figure}

Additionally atmospheric ozone profiles are seen in some observations. This has appeared in calibrated spectra due to atmospheric variation and receiver gain variation. The gain of P0 and P1 vary independently, so ozone spectra shape variation during an observation could be different by polarization when it is related to gain variation (see Figure~\ref{fig:ozone}). The procedure to confirm the ozone profile is introduced at EAO the website\footnote{\label{EAO_ozone}\url{https://www.eaobservatory.org/jcmt/instrumentation/heterodyne/ozone-lines/}}.

Issues often can be recognized by comparing two polarization data, so we are recommending  users reduce the polarizations separately to assess data quality \footnote{The method is summarized on the EAO website \url{https://www.eaobservatory.org/jcmt/instrumentation/heterodyne/data-reduction/reducing-acsis-data/\#Polarization-separated_reduction_Uu_specific}}.

\section{\texorpdfstring{$`$\=U$`$\=u: Science Demonstration}{}}

To test and demonstrate the ability of \Uu we provide observations of the Whirlpool Galaxy galaxy M51 and the most active star-forming region in the Orion Nebula, Orion KL (Kleinmann-Low Nebula; \cite{Becklin1967,Kleinmann1967}) 

\subsection{Galactic observations: Orion KL}

The increased mapping speed of \Uu in comparison to its predecessor is fundamentally important in the context of local star-forming regions. The right panel of Figure~\ref{fig:M51+Orion} shows the  CO (2-1) emission of the most active star-forming region in the Orion Nebula, Orion KL (Kleinmann-Low Nebula; \cite{Becklin1967,Kleinmann1967}) over 12 velocity channels. 

The $245''\times245''$ map was obtained over the course of \mbox{24 minutes} centered on position \mbox{R.A. = 5h35m14.5s}, \mbox{dec = -5$^{\circ}$:23':09.6''} in good weather conditions ($\tau_{225} = 0.08$). The spectral axis was binned such that the radio velocity resolution is \mbox{0.8 km/s}, the measured peak corrected antenna temperature, $T_{A}*$, is \mbox{67.5 K} with an RMS noise of $T_{A}*=0.2\mathrm{\:K}$. To obtain such a map with \Uu's predecessor RxA3m to the same depth would have required 1.7 hours of integration, 4.2 times longer.

\subsection{Galaxy Observations: M51}

\begin{figure}
  \centering
    \includegraphics[width=0.4\linewidth]{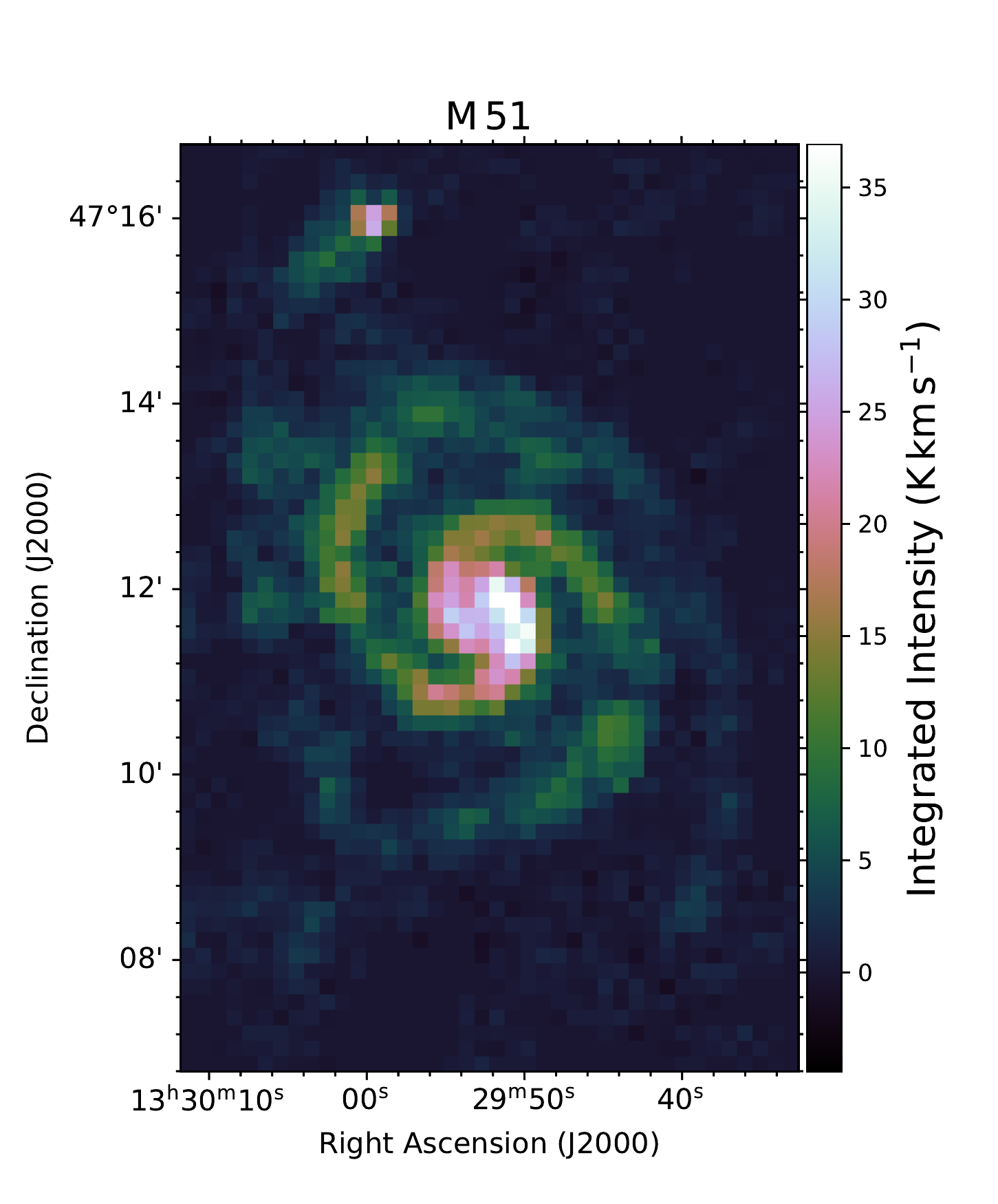}
  \includegraphics[width=0.55\linewidth]{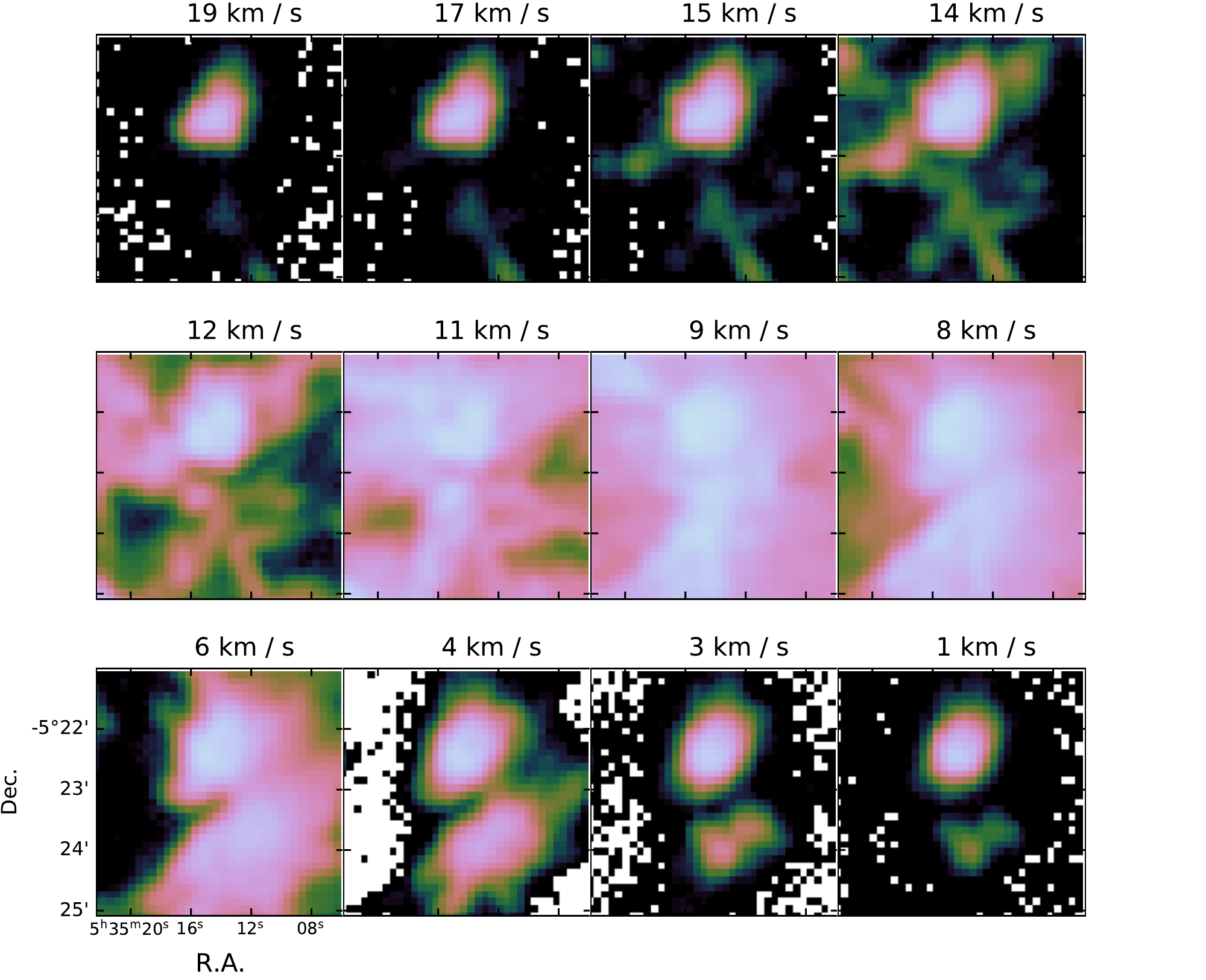}
  \caption{
  Left: CO (2-1) Integrated intensity map of M51 also known as the Whirlpool Galaxy, NGC5194. Right: CO (2-1) velocity channel maps of Orion KL, the most active star-forming region near the Earth. 
  }
  \label{fig:M51+Orion}
\end{figure}

From proposals already being observed at the JCMT under Shared Risk Observing it is known that astronomers performing galaxy studies will be one of the dominant science communities to make use of the \Uu receiver. In this section we present CO (2-1) \Uu observations of the Whirlpool Galaxy M\,51 (see left panel of Figure~\ref{fig:M51+Orion}. M\,51 (Messier 51) is a famous nearby interacting galaxy-pair, with the bigger one well-known as the Whirlpool galaxy. It has been extensively studied observationally at nearly all wavelengths. CO emission has been mapped for M\,51 in CO (1-0) \cite{Koda2011}, CO (2-1) \cite{Schuster2007} and CO (3-2) \cite{Vlahakis2013}. 

The \Uu mapping observations of CO (2-1) towards M\,51 in good weather conditions ($\tau_{225} \sim$ 0.07) took two hours to complete. The position-switch raster-scan mode map was centered on R.A. = 13:29:52.698, dec = 47:11:42.93 with a $600'' \times 400''$ area mapped, corresponding to the central $\sim$ 9 kpc of M\,51. Under the same conditions this observation (depth of $\sim 0.03$\,K at 12.7 km\,s$^{-1}$ resolution with measured peak corrected antenna temperature, $T_{A}*$, is 0.7\,K.) would have required $\sim$ 11.7 hours of integration time to reach a similar depth using the now retired RxA3m, $\sim$ 6 times longer. The integrated intensity (moment 0) map of M\,51 is shown in Figure~\ref{fig:M51+Orion}.

\section{summary}
The \Namakanui receiver, a spare receiver for the Greenland Telescope \cite{Han2018} with three inserts operating around 86, 230 and 345\,GHz, was installed at the JCMT in the summer of 2019. Basic functional check such as T\textsubscript{rx} and phase stabillity for all inserts have been performed. The on sky commissioning of \Uu is close to complete after a comprehensive review of pointing, beam size, calibration accuracy, leakage, polarization and sideband performance. The \Uu insert is currently available to the JCMT User community under Shared Risk Observing, with observations of the Orion Nebula and M51 demonstrating its functionality. 

Current observations provide evidence for \Uu's superior performance compared to the retired RxA3m, reaching similar depths at a rate 4-6 times faster than its predecessor. Considering such recent achievements from RxA3m such as the Event Horizon Telescope image of a Black Hole\cite{EHT2019}  and the detection of phosphine on Venus \cite{Greaves2020}, the future for this \Namakanui instrument is exciting. 

\acknowledgments     
 
The James Clerk Maxwell Telescope is operated by the East Asian Observatory on behalf of The National Astronomical Observatory of Japan; Academia Sinica Institute of Astronomy and Astrophysics; the Korea Astronomy and Space Science Institute; Center for Astronomical Mega-Science (as well as the National Key R$\&$D Program of China with No. 2017YFA0402700). Additional funding support is provided by the Science and Technology Facilities Council of the United Kingdom and participating universities in the United Kingdom and Canada. The James Clerk Maxwell Telescope has historically been operated by the Joint Astronomy Centre on behalf of the Science and Technology Facilities Council of the United Kingdom, the National Research Council of Canada and the Netherlands Organisation for Scientific Research. 

\Namakanui was constructed and funded by ASIAA in Taiwan, with funding for the mixers provided by ASIAA and at 230GHz by EAO. The \Namakanui instrument is a backup receiver for the GLT.

The authors wish to recognize and acknowledge the very significant cultural role and reverence that the summit of Maunakea has always had within the indigenous Hawaiian community.  We are most fortunate to have the opportunity to conduct astronomical observations from this mountain. Finally we wish to thank Dr. Larry Kimura for the Hawaiian naming of \Namakanui and it's three inserts: \Alaihi, \Uu and \Aweoweo. Dr. Kimura is an associate professor at University of Hawai‘i at Hilo Ka Haka ‘Ula o Ke‘elikolani College of Hawaiian Language. The name \Namakanui means "the big eyes", this name was recommended by Dr Kimura as big eyes being needed to see in the dark, much like \Namakanui observing the Universe. The three inserts \Alaihi, \Uu and \Aweoweo are the names of three big eyed nocturnal fish found in Hawai`i that have larger eyes to see in the dark, they are red in colour, as is the colour of the \Namakanui dewar.

\bibliography{report}   

\begin{thebibliography}{10}

\bibitem{Holland2013}
{Holland}, W.~S., {Bintley}, D., {Chapin}, E.~L., {Chrysostomou}, A., {Davis},
  G.~R., {Dempsey}, J.~T., {Duncan}, W.~D., {Fich}, M., {Friberg}, P.,
  {Halpern}, M., {Irwin}, K.~D., {Jenness}, T., {Kelly}, B.~D., {MacIntosh},
  M.~J., {Robson}, E.~I., {Scott}, D., {Ade}, P.~A.~R., {Atad-Ettedgui}, E.,
  {Berry}, D.~S., {Craig}, S.~C., {Gao}, X., {Gibb}, A.~G., {Hilton}, G.~C.,
  {Hollister}, M.~I., {Kycia}, J.~B., {Lunney}, D.~W., {McGregor}, H.,
  {Montgomery}, D., {Parkes}, W., {Tilanus}, R.~P.~J., {Ullom}, J.~N.,
  {Walther}, C.~A., {Walton}, A.~J., {Woodcraft}, A.~L., {Amiri}, M.,
  {Atkinson}, D., {Burger}, B., {Chuter}, T., {Coulson}, I.~M., {Doriese},
  W.~B., {Dunare}, C., {Economou}, F., {Niemack}, M.~D., {Parsons}, H.~A.~L.,
  {Reintsema}, C.~D., {Sibthorpe}, B., {Smail}, I., {Sudiwala}, R., and
  {Thomas}, H.~S., ``{SCUBA-2: the 10 000 pixel bolometer camera on the James
  Clerk Maxwell Telescope},'' {\em MNRAS}~{\bf 430}(4),  2513--2533 (2013).

\bibitem{Friberg2018}
{Friberg}, P., {Berry}, D., {Savini}, G., {Bintley}, D., {Dempsey}, J.,
  {Graves}, S., and {Parsons}, H., ``{Characterizing and reducing the POL-2
  instrumental polarization},'' in [{\em Millimeter, Submillimeter, and
  Far-Infrared Detectors and Instrumentation for Astronomy
  IX}{\nolinebreak\hspace{0.1em}]},  {Zmuidzinas}, J. and {Gao}, J.-R., eds.,
  {\em Society of Photo-Optical Instrumentation Engineers (SPIE) Conference
  Series} {\bf 10708},  107083M (2018).

\bibitem{Buckle2009}
{Buckle}, J.~V., {Hills}, R.~E., {Smith}, H., {Dent}, W.~R.~F., {Bell}, G.,
  {Curtis}, E.~I., {Dace}, R., {Gibson}, H., {Graves}, S.~F., {Leech}, J.,
  {Richer}, J.~S., {Williamson}, R., {Withington}, S., {Yassin}, G., {Bennett},
  R., {Hastings}, P., {Laidlaw}, I., {Lightfoot}, J.~F., {Burgess}, T.,
  {Dewdney}, P.~E., {Hovey}, G., {Willis}, A.~G., {Redman}, R., {Wooff}, B.,
  {Berry}, D.~S., {Cavanagh}, B., {Davis}, G.~R., {Dempsey}, J., {Friberg}, P.,
  {Jenness}, T., {Kackley}, R., {Rees}, N.~P., {Tilanus}, R., {Walther}, C.,
  {Zwart}, W., {Klapwijk}, T.~M., {Kroug}, M., and {Zijlstra}, T.,
  ``{HARP/ACSIS: a submillimetre spectral imaging system on the James Clerk
  Maxwell Telescope},'' {\em MNRAS}~{\bf 399}(2),  1026--1043 (2009).

\bibitem{EAVN2018}
{An}, T., {Sohn}, B.~W., and {Imai}, H., ``{Capabilities and prospects of the
  East Asia Very Long Baseline Interferometry Network},'' {\em Nature
  Astronomy}~{\bf 2},  118--125 (2018).

\bibitem{Han2018}
{Han}, C.-C., {Chen}, M.-T., {Huang}, Y.-D., {Kubo}, D., {Chang}, C.-C.,
  {Chang}, S.-H., {Wei}, T.-S., {Huang}, J.-D., {Chen}, C.-C., {Raffin}, P.,
  {Liu}, C.-T., {Ho}, P. T.~P., {Inoue}, M., {Matsushita}, S., {Asada}, K.,
  {Norton}, T.~J., {Chilson}, R., {Srinivasan}, R., {Liu}, K.-Y., {Li}, C.-T.,
  {Bintley}, D., {Walther}, C., {Friberg}, P., {Dempsey}, J., {Ogawa}, H.,
  {Kimura}, K., {Hasagawa}, Y., and {Srikanth}, S., ``{The first-light
  receivers for the Greenland Telescope},'' in [{\em Millimeter, Submillimeter,
  and Far-Infrared Detectors and Instrumentation for Astronomy
  IX}{\nolinebreak\hspace{0.1em}]},  {\em Society of Photo-Optical
  Instrumentation Engineers (SPIE) Conference Series} {\bf 10708},  1070835
  (2018).

\bibitem{Bintley2018}
{Bintley}, D., {Friberg}, P., {Berthold}, R., {Chuter}, T., {Liu}, K.-Y.,
  {Walther}, C., {Dempsey}, J., {Ho}, P., {McGregor}, H., {Matsushita}, S.,
  {Asada}, K., {Koyama}, S., {Nishioka}, H., {Han}, C.-C., {Huang}, C.-W.,
  {Lin}, L., {Wei}, T.-S., {Kubo}, D., {Srinivasan}, R., {Rao}, R., {Bower},
  G., {Oshiro}, P., and {Chen}, M.-T., ``{GLT receiver commissioning at JCMT
  and future JCMT instrumentation},'' in [{\em Millimeter, Submillimeter, and
  Far-Infrared Detectors and Instrumentation for Astronomy
  IX}{\nolinebreak\hspace{0.1em}]},  {Zmuidzinas}, J. and {Gao}, J.-R., eds.,
  {\em Society of Photo-Optical Instrumentation Engineers (SPIE) Conference
  Series} {\bf 10708},  1070815 (2018).

\bibitem{Ediss2004}
{Ediss}, G.~A., {Carter}, M., {Cheng}, J., {Effland}, J.~E., {Grammer}, W.,
  {Horner}, N., J., {Kerr}, A.~R., {Koller}, D., {Lauria}, E.~F., {Morris}, G.,
  {Pan}, S.~K., {Reiland}, G., and {Sullivan}, M., ``{ALMA Band 6 Cartridge:
  Design and Performance},'' in [{\em Fifteenth International Symposium on
  Space Terahertz Technology}{\nolinebreak\hspace{0.1em}]},  {Narayanan}, G.,
  ed.,  181--188 (2004).

\bibitem{Kerr2014}
{Kerr}, A.~R., {Pan}, S.-K., {Claude}, S. M.~X., {Dindo}, P., {Lichtenberger},
  A.~W., {Effland}, J.~E., and {Lauria}, E.~F., ``{Development of the ALMA
  Band-3 and Band-6 Sideband-Separating SIS Mixers},'' {\em IEEE Transactions
  on Terahertz Science and Technology}~{\bf 4}(2),  201--212 (2014).

\bibitem{Mahieu2012}
{Mahieu}, S., {Maier}, D., {Lazareff}, B., {Navarrini}, A., {Celestin}, G.,
  {Chalain}, J., {Geoffroy}, D., {Laslaz}, F., and {Perrin}, G., ``{The ALMA
  Band-7 Cartridge},'' {\em IEEE Transactions on Terahertz Science and
  Technology}~{\bf 2}(1),  29--39 (2012).

\bibitem{Dent-ACSIS-2000}
{Dent}, W., {Duncan}, W., {Ellis}, M., {Harris}, J., {Lightfoot}, J., {Wall},
  R., {Gibson}, H., {Hills}, R., {Richer}, J., {Smith}, H., {Withington}, S.,
  {Burgess}, T., {Casorso}, R., {Dewdney}, P., {Hovey}, G., {Redman}, R.,
  {Yeung}, K., {Force}, B., and {Pain}, I., ``{HARP and ACSIS on the JCMT},''
  in [{\em Imaging at Radio through Submillimeter
  Wavelengths}{\nolinebreak\hspace{0.1em}]},  {Mangum}, J.~G. and {Radford}, S.
  J.~E., eds., {\em Astronomical Society of the Pacific Conference Series} {\bf
  217},  33 (2000).

\bibitem{Casper2013}
Casper, {\em ROACH2} (2013).
\newblock https://casper.ssl.berkeley.edu/wiki/ROACH2.

\bibitem{MIT2020}
MIT, {\em Mark 6 VLBI Data Recording System} (2020).
\newblock https://www.haystack.mit.edu/tech/vlbi/mark6/index.html.

\bibitem{Warmels2020}
{Remijan}, A., {Biggs}, A., {Cortes}, P., {Dent}, B., {Di Francesco}, J.,
  {Fomalont}, E., {Hales}, A., {Kameno}, S., {Mason}, B., {Philips}, N.,
  {Saini}, K., {Stoehr}, F., {Vila Vilaro}, B., and {Villard}, E., {\em ALMA
  TechnicalHandbook, ALMA Doc. 8.3, ver. 1.0} (2020).
\newblock
  https://almascience.nao.ac.jp/documents-and-tools/cycle8/alma-technical-handbook.

\bibitem{2017Thompson}
{Thompson}, A.~R., {Moran}, J.~M., and {Swenson}, George~W., J.,  [{\em
  {Interferometry and Synthesis in Radio Astronomy, 3rd
  Edition}}{\nolinebreak\hspace{0.1em}]}, Springer (2017).

\bibitem{Finger2013}
{Finger}, R., {Mena}, P., {Reyes}, N., {Rodriguez}, R., and {Bronfman}, L.,
  ``{A Calibrated Digital Sideband Separating Spectrometer for Radio Astronomy
  Applications},'' {\em PASP}~{\bf 125}(925),  263 (2013).

\bibitem{Becklin1967}
{Becklin}, E.~E. and {Neugebauer}, G., ``{Observations of an Infrared Star in
  the Orion Nebula},'' {\em The Astrophysical Journal}~{\bf 147},  799 (1967).

\bibitem{Kleinmann1967}
{Kleinmann}, D.~E. and {Low}, F.~J., ``{Discovery of an Infrared Nebula in
  Orion},'' {\em The Astrophysical Journal Letters}~{\bf 149},  L1 (1967).

\bibitem{Koda2011}
{Koda}, J., {Sawada}, T., {Wright}, M. C.~H., {Teuben}, P., {Corder}, S.~A.,
  {Patience}, J., {Scoville}, N., {Donovan Meyer}, J., and {Egusa}, F., ``{CO(J
  = 1-0) Imaging of M51 with CARMA and the Nobeyama 45 m Telescope},'' {\em
  ApJS}~{\bf 193}(1),  19 (2011).

\bibitem{Schuster2007}
{Schuster}, K.~F., {Kramer}, C., {Hitschfeld}, M., {Garcia-Burillo}, S., and
  {Mookerjea}, B., ``{A complete $^{12}$CO 2-1 map of M 51 with HERA. I. Radial
  averages of CO, H{\,}I, and radio continuum},'' {\em A\&A}~{\bf 461}(1),
  143--151 (2007).

\bibitem{Vlahakis2013}
{Vlahakis}, C., {van der Werf}, P., {Israel}, F.~P., and {Tilanus}, R.~P.~J.,
  ``{A CO J = 3-2 map of M51 with HARP-B: radial properties of the spiral
  structure},'' {\em MNRAS}~{\bf 433}(3),  1837--1861 (2013).

\bibitem{EHT2019}
{Event Horizon Telescope Collaboration}, {Akiyama}, K., {Alberdi}, A., {Alef},
  W., {Asada}, K., {Azulay}, R., {Baczko}, A.-K., {Ball}, D., {Balokovi{\'c}},
  M., {Barrett}, J., {Bintley}, D., {Blackburn}, L., {Boland}, W., {Bouman},
  K.~L., {Bower}, G.~C., {Bremer}, M., {Brinkerink}, C.~D., {Brissenden}, R.,
  {Britzen}, S., {Broderick}, A.~E., {Broguiere}, D., {Bronzwaer}, T., {Byun},
  D.-Y., {Carlstrom}, J.~E., {Chael}, A., {Chan}, C.-k., {Chatterjee}, S.,
  {Chatterjee}, K., {Chen}, M.-T., {Chen}, Y., {Cho}, I., {Christian}, P.,
  {Conway}, J.~E., {Cordes}, J.~M., {Crew}, G.~B., {Cui}, Y., {Davelaar}, J.,
  {De Laurentis}, M., {Deane}, R., {Dempsey}, J., {Desvignes}, G., {Dexter},
  J., {Doeleman}, S.~S., {Eatough}, R.~P., {Falcke}, H., {Fish}, V.~L.,
  {Fomalont}, E., {Fraga-Encinas}, R., {Freeman}, W.~T., {Friberg}, P.,
  {Fromm}, C.~M., {G{\'o}mez}, J.~L., {Galison}, P., {Gammie}, C.~F.,
  {Garc{\'\i}a}, R., {Gentaz}, O., {Georgiev}, B., {Goddi}, C., {Gold}, R.,
  {Gu}, M., {Gurwell}, M., {Hada}, K., {Hecht}, M.~H., {Hesper}, R., {Ho},
  L.~C., {Ho}, P., {Honma}, M., {Huang}, C.-W.~L., {Huang}, L., {Hughes},
  D.~H., {Ikeda}, S., {Inoue}, M., {Issaoun}, S., {James}, D.~J., {Jannuzi},
  B.~T., {Janssen}, M., {Jeter}, B., {Jiang}, W., {Johnson}, M.~D., {Jorstad},
  S., {Jung}, T., {Karami}, M., {Karuppusamy}, R., {Kawashima}, T., {Keating},
  G.~K., {Kettenis}, M., {Kim}, J.-Y., {Kim}, J., {Kim}, J., {Kino}, M.,
  {Koay}, J.~Y., {Koch}, P.~M., {Koyama}, S., {Kramer}, M., {Kramer}, C.,
  {Krichbaum}, T.~P., {Kuo}, C.-Y., {Lauer}, T.~R., {Lee}, S.-S., {Li}, Y.-R.,
  {Li}, Z., {Lindqvist}, M., {Liu}, K., {Liuzzo}, E., {Lo}, W.-P., {Lobanov},
  A.~P., {Loinard}, L., {Lonsdale}, C., {Lu}, R.-S., {MacDonald}, N.~R., {Mao},
  J., {Markoff}, S., {Marrone}, D.~P., {Marscher}, A.~P., {Mart{\'\i}-Vidal},
  I., {Matsushita}, S., {Matthews}, L.~D., {Medeiros}, L., {Menten}, K.~M.,
  {Mizuno}, Y., {Mizuno}, I., {Moran}, J.~M., {Moriyama}, K., {Moscibrodzka},
  M., {M{\"u}ller}, C., {Nagai}, H., {Nagar}, N.~M., {Nakamura}, M., {Narayan},
  R., {Narayanan}, G., {Natarajan}, I., {Neri}, R., {Ni}, C., {Noutsos}, A.,
  {Okino}, H., {Olivares}, H., {Ortiz-Le{\'o}n}, G.~N., {Oyama}, T.,
  {{\"O}zel}, F., {Palumbo}, D. C.~M., {Patel}, N., {Pen}, U.-L., {Pesce},
  D.~W., {Pi{\'e}tu}, V., {Plambeck}, R., {PopStefanija}, A., {Porth}, O.,
  {Prather}, B., {Preciado-L{\'o}pez}, J.~A., {Psaltis}, D., {Pu}, H.-Y.,
  {Ramakrishnan}, V., {Rao}, R., {Rawlings}, M.~G., {Raymond}, A.~W.,
  {Rezzolla}, L., {Ripperda}, B., {Roelofs}, F., {Rogers}, A., {Ros}, E.,
  {Rose}, M., {Roshanineshat}, A., {Rottmann}, H., {Roy}, A.~L., {Ruszczyk},
  C., {Ryan}, B.~R., {Rygl}, K. L.~J., {S{\'a}nchez}, S.,
  {S{\'a}nchez-Arguelles}, D., {Sasada}, M., {Savolainen}, T., {Schloerb},
  F.~P., {Schuster}, K.-F., {Shao}, L., {Shen}, Z., {Small}, D., {Sohn}, B.~W.,
  {SooHoo}, J., {Tazaki}, F., {Tiede}, P., {Tilanus}, R. P.~J., {Titus}, M.,
  {Toma}, K., {Torne}, P., {Trent}, T., {Trippe}, S., {Tsuda}, S., {van
  Bemmel}, I., {van Langevelde}, H.~J., {van Rossum}, D.~R., {Wagner}, J.,
  {Wardle}, J., {Weintroub}, J., {Wex}, N., {Wharton}, R., {Wielgus}, M.,
  {Wong}, G.~N., {Wu}, Q., {Young}, K., {Young}, A., {Younsi}, Z., {Yuan}, F.,
  {Yuan}, Y.-F., {Zensus}, J.~A., {Zhao}, G., {Zhao}, S.-S., {Zhu}, Z.,
  {Algaba}, J.-C., {Allardi}, A., {Amestica}, R., {Anczarski}, J., {Bach}, U.,
  {Baganoff}, F.~K., {Beaudoin}, C., {Benson}, B.~A., {Berthold}, R.,
  {Blanchard}, J.~M., {Blundell}, R., {Bustamente}, S., {Cappallo}, R.,
  {Castillo-Dom{\'\i}nguez}, E., {Chang}, C.-C., {Chang}, S.-H., {Chang},
  S.-C., {Chen}, C.-C., {Chilson}, R., {Chuter}, T.~C., {C{\'o}rdova Rosado},
  R., {Coulson}, I.~M., {Crawford}, T.~M., {Crowley}, J., {David}, J.,
  {Derome}, M., {Dexter}, M., {Dornbusch}, S., {Dudevoir}, K.~A., {Dzib},
  S.~A., {Eckart}, A., {Eckert}, C., {Erickson}, N.~R., {Everett}, W.~B.,
  {Faber}, A., {Farah}, J.~R., {Fath}, V., {Folkers}, T.~W., {Forbes}, D.~C.,
  {Freund}, R., {G{\'o}mez-Ruiz}, A.~I., {Gale}, D.~M., {Gao}, F., {Geertsema},
  G., {Graham}, D.~A., {Greer}, C.~H., {Grosslein}, R., {Gueth}, F., {Haggard},
  D., {Halverson}, N.~W., {Han}, C.-C., {Han}, K.-C., {Hao}, J., {Hasegawa},
  Y., {Henning}, J.~W., {Hern{\'a}ndez-G{\'o}mez}, A., {Herrero-Illana}, R.,
  {Heyminck}, S., {Hirota}, A., {Hoge}, J., {Huang}, Y.-D., {Impellizzeri},
  C.~M.~V., {Jiang}, H., {Kamble}, A., {Keisler}, R., {Kimura}, K., {Kono}, Y.,
  {Kubo}, D., {Kuroda}, J., {Lacasse}, R., {Laing}, R.~A., {Leitch}, E.~M.,
  {Li}, C.-T., {Lin}, L. C.~C., {Liu}, C.-T., {Liu}, K.-Y., {Lu}, L.-M.,
  {Marson}, R.~G., {Martin-Cocher}, P.~L., {Massingill}, K.~D., {Matulonis},
  C., {McColl}, M.~P., {McWhirter}, S.~R., {Messias}, H., {Meyer-Zhao}, Z.,
  {Michalik}, D., {Monta{\~n}a}, A., {Montgomerie}, W., {Mora-Klein}, M.,
  {Muders}, D., {Nadolski}, A., {Navarro}, S., {Neilsen}, J., {Nguyen}, C.~H.,
  {Nishioka}, H., {Norton}, T., {Nowak}, M.~A., {Nystrom}, G., {Ogawa}, H.,
  {Oshiro}, P., {Oyama}, T., {Parsons}, H., {Paine}, S.~N., {Pe{\~n}alver}, J.,
  {Phillips}, N.~M., {Poirier}, M., {Pradel}, N., {Primiani}, R.~A., {Raffin},
  P.~A., {Rahlin}, A.~S., {Reiland}, G., {Risacher}, C., {Ruiz}, I.,
  {S{\'a}ez-Mada{\'\i}n}, A.~F., {Sassella}, R., {Schellart}, P., {Shaw}, P.,
  {Silva}, K.~M., {Shiokawa}, H., {Smith}, D.~R., {Snow}, W., {Souccar}, K.,
  {Sousa}, D., {Sridharan}, T.~K., {Srinivasan}, R., {Stahm}, W., {Stark},
  A.~A., {Story}, K., {Timmer}, S.~T., {Vertatschitsch}, L., {Walther}, C.,
  {Wei}, T.-S., {Whitehorn}, N., {Whitney}, A.~R., {Woody}, D.~P.,
  {Wouterloot}, J. G.~A., {Wright}, M., {Yamaguchi}, P., {Yu}, C.-Y.,
  {Zeballos}, M., {Zhang}, S., and {Ziurys}, L., ``{First M87 Event Horizon
  Telescope Results. I. The Shadow of the Supermassive Black Hole},'' {\em
  ApJL}~{\bf 875}(1),  L1 (2019).

\bibitem{Greaves2020}
{Greaves}, J.~S., {Richards}, A. M.~S., {Bains}, W., {Rimmer}, P.~B., {Sagawa},
  H., {Clements}, D.~L., {Seager}, S., {Petkowski}, J.~J., {Sousa-Silva}, C.,
  {Ranjan}, S., {Drabek-Maunder}, E., {Fraser}, H.~J., {Cartwright}, A.,
  {Mueller-Wodarg}, I., {Zhan}, Z., {Friberg}, P., {Coulson}, I., {Lee}, E.,
  and {Hoge}, J., ``{Phosphine gas in the cloud decks of Venus},'' {\em Nature
  Astronomy}  (2020).

\end{thebibliography}
\bibliographystyle{spiebib}   

\end{document}